
\magnification=\magstep1
\baselineskip=24truept
\pageno=1

\centerline {\bf Uncertainty Relation at Finite Temperature}
\vskip .5cm
\centerline{\bf B. L. Hu $^{\ast}$}
\centerline{\it Department of Physics, University of Maryland,
College Park, MD 20742, USA}
\vskip .5cm
\centerline{\bf Yuhong Zhang ${\ast}$}
\centerline{\it Biophysics Laboratory,
Center for Biologics Evaluation and Research, }
\centerline{\it Food and Drug Adminstration, 8800 Rockville Pike, Bethesda,
MD 20892, USA}
\vskip .5cm
\centerline{(umdpp 93-161)}
\vskip .5cm

\centerline{\bf Abstract}

\vskip .2 cm

We use the quantum Brownian model
to derive the uncertainty relation for a quantum open system in an
arbitrarily-squeezed initial state
interacting with an environment at finite temperature. We examine
the relative importance of the quantum and thermal fluctuations in the
evolution
of the system towards equilibrium with the aim of clarifying the
meaning of quantum, classical and thermal.
We show that upon contact with the bath
the system evolves from a quantum-dominated
state to a thermal-dominated state in a time which is the same as
the decoherence time calculated before in the context of quantum to classical
transitions. We also use these results to deduce the conditions
when the two basic postulates of quantum statistical mechanics become valid.

 \vskip 2cm
\noindent - Revised version submitted to Physical Review Letters, July 30, 1993
\noindent $^{\ast}$ e-mail addresses: hu@umdhep.bitnet, zhang@helix.nih.gov

\vfill
\eject

It is a well-known fact in quantum mechanics that a lower bound exists in
the product of the variances of pairs of noncommutative observables.
Taking the coordinate $x$ and momentum $p$ as examples, the Heisenberg
uncertainty principle states that with $(\Delta x)^2=~<x^2>-~<x>^2$,
and  $(\Delta p)^2=~<p^2>-~<p>^2$, the uncertainty function
$$
U_0 ^{QM} =(\Delta x)^2(\Delta p)^2 \geq {\hbar^2 \over 4}
{}~~~(T=0,~~ quantum~ mechanics)
                                                              \eqno(1)
$$
The existence of quantum fluctuations is a verified basic physical
phenomenon. The origin of the uncertainty relation can be attributed as
a mathematical property of Fourier analysis [1] which describes quantum
mechanics as a wave theory. Recent years have seen effort in establishing
a stronger relation based on information-theoretical considerations [2][3].

In realistic conditions quantum systems are often prepared and studied
at finite temperatures where thermal fluctuations permeate. At high
temperatures the equipartition theorem of classical statistical mechanics
imparts for each degree of freedom an uncertainty of $kT/2$.
Thus the uncertainty function for a one-dimensional harmonic oscillator
approaches the limit
$$
U_T ^{MB} \approx ({kT\over\Omega_0})^2
{}~~~(high~T,~~ classical~ statistical~ mechanics)
                                                                \eqno(2)
$$
where $\hbar \Omega_0$ is the energy of a normal mode with natural
frequency $\Omega_0$.
This result, obtained by assuming that the system obeys the
Maxwell-Boltzmann distribution, is  usually regarded as the classical limit.
For a system of bosons in equilibrium at temperature $T$, the application
of canonical ensemble gives the result in quantum statistical mechanics
as
$$
U_T ^{BE} = {\hbar^2 \over 4} [ \coth ({{ \hbar \Omega_0} \over {2 kT}})]^{2}
{}~~~(all~ T,~~quantum~ statistical~ mechanics)
                                                                \eqno(3)
$$
which interpolates between the two results (1) and (2) at $T=0$ and
$T>> \hbar \Omega_0/k$. This result applies to a system already in
equilibrium at temperature $T$.

Our purpose here is to study the corresponding non-equilibrium problem.
At time $t_0$ we put the system in contact with a heat bath at temperature
$T$ and follow its time evolution. We want to see how the uncertainty
function $U_T(t)$ changes from the initial quantum fluctuation-dominated
condition to a later thermal fluctuation-dominated condition. By comparing
this result with the decoherence studies recently carried out [4,5,6],
where two characteristic times---the decoherence time $t_{dec}$ and the
relaxation time $t_{rel}$---are defined, one can apply the physics of
these two processes involved to examine the following three interrelated
issues :

\noindent
1) the realization of the basic tenets of quantum statistical
mechanics from quantum dynamics;
2) the relation between quantum and thermal fluctuations and
3) their role in the quantum to classical transition.

Quantum statistical mechanics of a macroscopic system is derived from
the quantum dynamics of its microscopic constituents under two basic
postulates [7]: i) random phase, and ii) equal {\it a priori} probability.
The first condition enables one to assign probability distributions to
a system occupying certain quantum states. It requires sufficient diminishing
of interference terms in the wave function or that the density matrix
of the system be approximately diagonal. The second condition when applied
to a closed system forms the basis of the microcanonical ensemble. When
applied to an open system in contact with a heat bath
described by the canonical ensemble, it ensures that the system
equilibrates with
its environment. We want to examine the processes by which these two conditions
are
attained from a more primitive level, starting with the microdynamics of
a system of quantum particles. Specifically, we want to see if there is
a characteristic time  when the phase information is lost (Postulate i)
and another time when the system attains equilibrium with its surrounding
so that all accesible states in the closed system are equally probable
(Postulate ii).

On the second issue, the demarkation of classical, quantum and thermal regimes
are not always clearly noted, their usual definitions or usage oftentimes
are imprecise. With the aid of the results obtained here we will point out
some existing confusions and make their meanings more precise. The relation
between quantum and thermal fluctuations has been studied previously via thermo
field dynamics [8] under equilibrium and stationary conditions. Here we
treat the fully nonequilibrium problem.
Using the microdynamics of a quantum system as
starting point we view thermal fluctuations as statistical variations of the
coarse-grained environmental variables with which the quantum system interacts,
the exact microdynamics of the system and the environment obeying the laws
of quantum mechanics. The problem under study can thus be stated equivalently
as finding the uncertainty relation for an open quantum system [9].

On the third issue, in loose terms, one often identifies the high temperature
regime of a system as the classical domain. On the one hand, one regards
the regime when thermal fluctuations begin to surpass quantum fluctuations
as the transition point from quantum to classical. On the other hand, from the
wave picture (or consistent history formulation) of quantum mechanics we know
that a necessary condition for a
system to behave classically is that the interference terms in its wave
function (or interference in the histories) have to diminish sufficiently and
remain insignificant
%
%
so that probability can be assigned to classical events [4,5]
%
%
or that classical decoherent histories can be well-defined [6]. This is
known as the decoherence process.
Is there any relation between these two criteria of
classicality? We show in this problem under the conditions studied that they
are indeed equivalent: The time the quantum system decoheres is also the time
when thermal fluctuation overtakes quantum fluctuations [see Eq. (22, 25)].

However we issue a warning here that this regime should not be called
classical,
as is customary in many quantum to classical transition studies. In fact,
after the decoherence time only the first postulate of quantum statistical
mechanics (QSM) is satisfied, the system can be described by {\it
non-equilibrium}
QSM. Only after the relaxation time, when the
second postulate (more precisely, the implied version for canonical ensemble)
is satisfied for the closed system,
can one use {\it equilibrium} QSM. Classical has still a long way to go.
It is well-known that quantum statistical effects can be important at very high
temperatures (e.g., Fermi temperature for metals). This is due to exchange
interactions of identical particles, a distinctly quantum effect. Only when
the statistical properties of fermions and bosons can be approximated by
distinguishable particles, usually at high temperatures when the Fermi-Dirac
or Bose-Einstein statistics approaches the Maxwell-Boltzmann statistics, can
the system be rightfully called {\it classical}. In this regard {\it quantum}
carries two meanings, one refers to the interference effect and the other
refers
to spin-statistics effect.

To examine these issues, we use the simple Brownian model [4-6, 10-13]
of a quantum open system and the influence functional
method [11] to incorporate the statistical effect of the environment
on the system.
As the microdynamics is explicitly treated in this approach,
one can study how the result
depends on the properties of the bath and the system-bath interaction in
detail.
We consider an initial Gaussian wave packet  (12) with nonzero averaged
initial position and momentum and calculate the time
dependence of the spread
$<\Delta x>^2, < \Delta p>^2 $ and the uncertainty function $U_T$ in the
presence of both quantum and thermal fluctuations. The main result is given
formally by Eq. (17), which, for an ohmic environment
under the conditions of weak damping and high temperature, simplifies to
Eqs. (19) and Eq. (24) respectively. As the initial state is
arbitrarily- squeezed, we can show clearly how the squeeze parameter
enters in these uncertainty relations [14].

Our system is a Brownian particle with mass $M$ and natural frequency
$\Omega_0$.
The environment is modeled by a set of $n$ harmonic oscillators with mass $m_n$
and natural frequency $\omega_n$. The particle is coupled linearly to the $n$th
oscillator with strength $C_n$. The action of the combined system and
environment is
$$
S[x,q] = \int\limits_0^tds\Biggl[
    \Bigl\{{1\over 2}M\dot x^2-{1\over 2}M\Omega_0^2 x^2\Bigr\}
  + \sum_n\Bigl\{{1\over 2}m_n\dot q_n^2
  - {1\over 2}m_n\omega^2_nq_n^2 \Bigr\}
  + \sum_n\Bigl\{-C_nxq_n\Bigr\}\Biggr]                          \eqno(4)
$$
\noindent where $x$ and $q_n$ are the coordinates of the particle and the
oscillators respectively, and $\Omega_0$ is the (bare) frequency of the
particle. We are interested in how the environment affects the system in
some averaged way.  The quantity containing this information is the reduced
density matrix of the system $\rho_r(x,x')$ obtained from the full density
operator of the system and environment $\rho(x,q;x'q')$ by tracing out the
environmental degrees of freedom $ (q,q') $
$$
\rho_r(x,x',t)
=\int\limits_{-\infty}^{+\infty}dq
 \int\limits_{-\infty}^{+\infty}dq'
 \rho(x,q;x',q',t)\delta(q-q')                    \eqno(5)
$$
\noindent The reduced density matrix evolves under the action of the
propagator $J_r(x,x',t~|~x_i,x'_i,0) $ in the following way:
$$
\rho_r(x,x',t)
=\int\limits_{-\infty}^{+\infty}dx_i
 \int\limits_{-\infty}^{+\infty}dx'_i~
 J_r(x,x',t~|~x_i,x'_i,0)~\rho_r(x_i,x'_i,0~)   \eqno(6)
$$
In general this is a very complicated expression since the evolution
operator $J_r$ depends on the initial state. If we assume that at a given
time $t=0$ the system and the environment are uncorrelated, i.e. that
$  \hat\rho(t=0)=\hat\rho_s\times\hat\rho_e, $ and that
the environment is in thermal equilibrium at a temperature
$T$,  the evolution operator for the reduced density matrix has an analytic
form
given by [5,11-13]: (We use the notation of [5])
$$
\eqalign{
J_r(x_f,x'_f,t~|~x_i,x'_i,0)
& = Z_0(t)  \exp{i\over\hbar} \Bigl\{
   \Bigl[\dot u_1(0)X_i+\dot u_2(0)X_f\Bigr]Y_i
  -\Bigl[\dot u_1(t)X_i+\dot u_2(t)X_f\Bigr]Y_f\Bigr\} \cr
& \times\exp{-1\over\hbar}\Bigl\{a_{11}(t)Y_i^2
  +[a_{12}(t)+a_{21}(t)]Y_iY_f
  +a_{22}(t)Y_f^2\Bigr\} \cr }                        \eqno(7)
$$
\noindent where $ X=(x+x')/2 $ and $Y=x'-x$. The elementary functions
$ u_a(s) $ are obtained as solutions of the
following differential equations
$$
{d^2u_i(s)\over ds^2}
+2\int\limits_0^sds'\eta(s-s')u_i(s')
+\Omega_0^2u_i(s)=0                                  \eqno(8)
$$
\noindent with boundary conditions
$$
\Biggl\{
\eqalign{
& u_1(0)=1, ~~~  u_1(t)= 0 \cr
& u_2(0)=0, ~~~  u_2(t)= 1 \cr }
    						      \eqno(9)
$$
\noindent and the $ a_{ij}(t) $ are obtained from the integral
$$
a_{ij}(t)
=\int\limits_0^tds_1\int\limits_0^{s_1}ds_2
 v_i(s_1)\nu(s_1-s_2)v_j(s_2)                       \eqno(10)
$$
\noindent where $ v_1(s)=u_2(t-s) $ and $ v_2(s)=u_1(t-s)$.
Here $ \eta $ and $ \nu $ are the dissipation and
noise kernels in the  influence functional [11].
They are given respectively by
$$
\eqalignno{
\nu(s)&
 =\int\limits_0^{+\infty}d\omega~I(\omega)
  \coth({{\hbar \omega} \over {2 k T}})  ~
  \cos\omega s                                      &(11a)\cr
\noalign{\hbox{and}}
\eta(s)&={d\over ds}~\gamma(s);~~
\gamma(s)=\int\limits_0^{+\infty}
d\omega~{I(\omega)\over\omega}~\cos\omega s.         &(11b)}
$$
\noindent Here $I(\omega)$ is the spectral density function of the environment,
$ I(\omega)= \sum\limits_n{\delta(\omega -\omega_n)}
           {{C^2_n}\over{2 m_n \omega_n}}.   $
An environment is called ohmic $I(\omega)\sim\omega$, supra-ohmic
$I(\omega)\sim\omega^n , n>1$ or sub-ohmic $n<1$. The most studied ohmic
case corresponds to an environment which induces a dissipative force
linear in the velocity of the system.

We now consider a Brownian oscillator with an initial wave function
$$
\psi(x,0)=
\sqrt{N_0}~\exp\Bigl\{-{(x-x_0)^2\over 4\sigma^2}
+{i\over\hbar}p_0 x \Bigr\}
                                                      \eqno(12)
$$

\noindent where $\sigma$ is the initial spread, and $x_0, p_0$ are the
averaged initial position and momentum of the Gaussian wave packet.
This gives $\rho_r (x, x', 0) = \psi ^\dagger (x, 0) \psi (x', 0)$.
One can calculate $\rho_r(x_f, x'_f, t)$ by performing the Gaussian integrals
over $x_i$ and $x'_i$ and get
$$
\rho_r(x_f, x'_f, t)
=\tilde Z_0 (t) \exp\Bigl\{-{1\over 2} {\bf X}^T {\bf Q}{\bf X}
                           + {\bf B}^T {\bf X}                    \Bigr\}
                                                   \eqno(13)
$$
where the prefactor $\tilde Z_0(t)$ depends only on time.
Here ${\bf X}=(X_f,Y_f)^T$. The parameters $x_0, p_0$ enter in the
elements $B_i$ of ${\bf B}$ but do not appear in the uncertainty function.
The elements $Q_{ij}(t), i, j= 1,2$ of ${\bf Q}$ are given below.

To calculate the averages of observables, it is convenient to use the
Wigner function defined as
$$
W_r(X,p,t) = \int dY e^{{i\over \hbar} p Y} \rho_r (X- {Y \over 2},
             X+ {Y \over 2}, t),                         \eqno(14)
$$
The quantum statistical average of a system observable,
e.g., $x^n$ or $p^n$, with respect to the reduced density matrix $\rho_r (t)$
is given by
$$
< x^n>_T = \int dx x^n \rho_r(x,x,t)
         = \int dx \int {{dp} \over {2 \pi \hbar}} x^n W_r(X, p, t)
                                                        \eqno(15)
$$
where the subscript $T$ indicates that the environment is at temperature $T$.
Obviously the average contains both quantum and thermal contributions.
We get
the variances
$$
 (\Delta x)_T^2 \equiv <x^2>_T-<x>_T^2 = {1\over Q_{11}(t)} ; ~~
 (\Delta p)_T^2 \equiv <p^2>_T-<p>_T^2 = \hbar^2 {\det {\bf Q}\over Q_{11}(t)}
                                                             \eqno(16)
$$

and arrive at the finite temperature uncertainty function:
$$
U_T (t) = (\Delta x)_T^2 (\Delta p)_T^2
        = \hbar^2 {\det {\bf Q}(t) \over [Q_{11}(t)]^2 }
        = \hbar^2 \Bigl\{ {Q_{22}(t) \over Q_{11}(t)}
        - {Q_{12}(t) \over Q_{11}(t)}
          {Q_{21}(t) \over Q_{11}(t)} \Bigr\}.
                                                        \eqno(17)
$$

\noindent The matrix elements in (17) are given by
$$
\eqalign{
{Q_{22}(t) \over Q_{11}(t)}
& = {1\over 4} { [\dot u_1(t)]^2 \over [\dot u_2(0)]^2 }
+ {2\sigma^2\over\hbar [\dot u_2(0)]^2} \Bigl\{
   [\dot u_1(t)]^2 a_{11}(t)
- \dot u_1(0) \dot u_1(t) [a_{12}(t)+a_{21}(t)] \cr
& + \Bigl[ {\hbar^2\over 4\sigma^4}
+ [\dot u_1(0)]^2\Bigr] a_{22}(t) \Bigr\}
+ {1\over [\dot u_2(0)]^2} \Bigl\{
  4a_{11}(t)a_{22}(t)-[a_{12}(t)+a_{21}(t)]^2 \Bigr\} \cr }
                                                        \eqno(18a)
$$
\noindent and
$$
\eqalign{
{Q_{12}(t) \over Q_{11}(t)}
= {Q_{21}(t) \over Q_{11}(t)}
& = \sigma^2 {i\dot u_2(t)\over \hbar [\dot u_2(0)]^2}
\Bigl\{ {\hbar^2\over 4\sigma^4} + [\dot u_1(0)]^2
-{\dot u_2(0)\over \dot u_2(t)} \dot u_1(0) \dot u_1(t)\Bigl\} \cr
& +{i\over \dot u_2(0)} \Bigl\{ 2 {\dot u_2(t) \over \dot u_2(0)}
a_{11}(t) + a_{12}(t) + a_{21}(t) \Bigr\} \cr }
                                                        \eqno(18b)
$$
The complete result can be obtained from solving the
equations for the $u_i(s)$ functions and the $a_{ij}(t)$ coefficients.
It is also instructional to look at some special
cases where one can find simpler analytical expressions.

For an ohmic environment
$ \gamma (t) = 2 \gamma_0 \delta (t) $ and $a_{ij}(t)$, $u_i(t)$ become
harmonic and exponential functions. Let us introduce the following
physical parameters:
the effective frequency
$\Omega \equiv (\Omega_0^2 - \gamma_0^2/4)^{1/2}$,
the energy parameter $\epsilon \equiv { \hbar \Omega \over 2kT}
\equiv 1/\tau$, the damping parameter $\alpha \equiv {\gamma_0\over 2 \Omega}
$ and the squeeze parameter $\delta \equiv {2\Omega_0\sigma^2 \over \hbar}$
which measures the spread in the initial Gaussian wavepacket.
For ohmic {\it weak damping}, $\alpha << 1$, $\Omega \simeq \Omega_0$
the uncertainty function is given {\it for all temperatures} to first order in
$\alpha$ by
$$
\eqalign{
{1\over\hbar^2}U_T(t)
& \sim {1 \over 4}\biggl\{ e^{-\gamma_0 t}
    +\coth\epsilon
     \Bigl[1-e^{-\gamma_0 t}\Bigr] \biggl\}^2                   \cr
& + \coth\epsilon
    \biggl\{ {(1-\delta)^2\over 4\delta}
    \Bigl[1-e^{-\gamma_0 t}\Bigr]
  -  {{1-\delta^2}\over 4\delta}
    \alpha~ \sin2\Omega t~  \biggr\}e^{-\gamma_0 t}          \cr
& +  \biggl\{ {{1-\delta^2}\over 4\delta} \sin 2\Omega t
  +  \alpha \Bigl[\coth\epsilon
    -{{1+\delta^2} \over 2\delta} \Bigr]
     \sin^2\Omega t \biggr\}^2 e^{-2\gamma_0 t}               \cr}
                                                        \eqno(19)
$$
\noindent We see that
there are two factors at play here: time and temperature. Time is measured
in units of the relaxation time proportional to $t_{rel}=\gamma_0^{-1}$, and
temperature is measured with reference to the ground state energy
$\hbar \Omega_0 /2$ of the system. There are also two parameters involved,
$\alpha$ and $\delta$.
One can deduce various limits from this expression.
For example, it is obvious that at $t=0$,
when the initial uncorrelated conditions is assumed valid,
$ U_T(0)=\hbar^2/4$, which is the Heisenberg relation (1).
At very long time ($t>>\gamma_0^{-1}$), $ U_T(t) $ is insensitive to  $\delta$
and approaches $ U^{BE}_T $ as in (3) at finite temperature (as $\Omega
\simeq \Omega_0$).
That means the Brownian particle approaches an equilibrium
quantum statistical system. (For supraohmic bath this may not always be true).
We can also see that for a  $T=0$ bath ($\coth \epsilon =1$),
$U_T(t)$ has a leading term given
by $\hbar^2/4$ (the Heisenberg relation) followed, for squeezed states
$\delta \neq 1$, by terms of order $\alpha^0$ and
$\alpha$ depicting both decay and oscillatory behavior.
We understand that this is due to the action of quantum fluctuations alone.
For a minimum- uncertainty initial state ($\delta=1$),
we get for all finite temperatures
$$
U_T(t)
={\hbar^2\over 4}\biggl\{ e^{-\gamma_0 t}
+ \coth\epsilon
 \Bigl[1-e^{-\gamma_0 t}\Bigr] \biggl\}^2 + O (\alpha^2)
                                                        \eqno(20)
$$
Notice that there is no linear order damping term.

At short times ($t<< \gamma_0^{-1}$),
$$
U_T(t)
\simeq {\hbar^2\over 4} \Bigl[ 1 + 2(\delta \coth {\hbar \Omega \over
2kT} - 1) \gamma_0 t + O (t^2) \Bigr]
                                                        \eqno(21)
$$
This simple  expression is revealing in several aspects:
The first term is the ubiquitous
quantum fluctuation, the second term is the thermal contribution, which depends
on the initial spread and increases with increasing dissipation and
temperature.
The time when thermal fluctuations overtake quantum fluctuations is
(assuming the temperature is higher than the ground state energy):
$$
t_1 = {1 \over {2 \gamma_0 (\delta \coth { \hbar \Omega \over 2kT} - 1)}}
\eqno(22)
$$
\noindent We expect this to be equal to  the decoherence time scale $t_{dec}$
calculated for weak coupling for all temperatures in the study
of quantum to classical transitions [15].

At zero coupling, i.e., for an isolated harmonic oscillator,
$$
U_T(t)
= {1\over 4} \hbar^2
\Bigl\{ 1 + { (1-\delta^2)^2 \over 4\delta^2 }
\sin^22\Omega_0t \Bigr\}
\ge {1\over 4} \hbar^2
                                                          \eqno(23)
$$
\noindent This is the quantum uncertainty relation for squeezed states.
The time-dependent term is the result of quantum dispersion.

At {\it high temperatures}, $\tau_0 \equiv {kT \over 2\hbar \Omega_0} >> 1$,
both the noise
kernel and the dissipation kernel for ohmic dissipation become local,
we have for the {\it full range of damping strength},
$$
\eqalign{
{1\over\hbar^2}U_T(t)
& = {1 \over 4}\biggl\{e^{-\gamma_0 t}
    + \tau_0 \Bigl[1-e^{-\gamma_0 t}\Bigr] \biggl\}^2             \cr
& + \tau_0 \biggl\{ {(1-\delta)^2\over 4\delta}
    \Bigl[1-e^{-\gamma_0 t}\Bigr]
    -\alpha  \Bigl[ {{1-\delta^2}\over 4\delta} \Bigr]\sin 2\Omega t
    -\alpha^2 \Bigl[ \tau_0
    -{{1+\delta^2} \over 2\delta} \Bigr]
     \sin^2\Omega t \biggr\} e^{-\gamma_0 t}                  \cr
& + {\Omega_0^2\over\Omega^2} \biggl\{ \Bigl[
     {{1-\delta^2}\over 4\delta} \Bigr] \sin 2\Omega t
  +  \alpha \Bigl[ \tau_0 -{{1+\delta^2} \over 2\delta} \Bigr]
     \sin^2\Omega t \biggr\}^2 e^{-2\gamma_0 t}               \cr}
                                                        \eqno(24)
$$
At short times ($t<< \gamma_0^{-1}$),
$$
U_T(t)
\simeq {\hbar^2\over 4} \Bigl[ 1 + 2(\tau_0\delta - 1) \gamma_0 t
+ O (t^2) \Bigr]
                                                        \eqno(25)
$$
The time when thermal fluctuations overtake quantum fluctuations is
given by
$$
t'_1 = {{ \hbar \Omega_0} \over {4 kT \gamma_0 \delta}}     \eqno(26)
$$
\noindent This is equal to  the decoherence time scale $t_{dec}$
calculated before for high temperature for the same model [4,5].

To complete the description of the evolution of the system,
the second time scale of importance is the relaxation time scale,
$t_{rel} = \gamma_0^{-1}$ we referred to earlier, when the particle reaches
equilibrium with the environment.
It is at this time that
equilibrium QSM can be applied to the description of the system.
After this, for ohmic and subohmic
environments the uncertainty relation takes on the Bose-Einstein form (3).
At high temperatures the system reaches the Maxwell-Boltzmann limit and the
uncertainty relation takes on the classical form (2).
For supraohmic environments at low temperature,
the highly nonlocal frequency response may
make it difficult for the system to settle down. The decoherence time scale is
longer, and the relaxation can even be incomplete.
This is the regime where one expects to find more intricate and interesting
behavior in the interplay of quantum and thermal effects. The nonohmic results
[16] and details of the present study [9] are to be presented elsewhere.


After this paper was submitted for publication we learned of the
work of Anderson and Halliwell [3] on an information-theoretical definition of
uncertainty which, among reporting other interesting results, also
confirmed our main findings here.

We thank Juan Pablo Paz and Alpan Raval for discussions, and
Jonathan Halliwell for explaining their information-theoretic results.
This work is supported in part by the National Science Foundation under
grant PHY 91-19726.

\vskip 1cm

\centerline{\bf References}

\noindent [1] I. Bialynicki-Birula and J. Mycielski, Comm. Math. Phys. 44,
    129 (1975);     W. Beckner, Ann. Math. 102, 159 (1975).

\noindent [2] D. Deutsch, Phys. Rev. Lett. 50, 631 (1983);
    M. H. Partovi, Phys. Rev. Lett. 50, 1883 (1983).
    See also S.Abe and N. Suzuki, Phys. Rev. A41, 4608 (1990).

\noindent [3] A. Anderson and J. J. Halliwell, Phys. Rev. D48 (1993);
              J. J. Halliwell, Imperial Collega Preprint IC92-93/26 (1993).

\noindent [4] W. H. Zurek, Phys. Rev. D24, 1516 (1981); D26, 1862 (1982);
    in {\it Frontiers of Nonequilibrium Statistical Physics},
    ed. G. T. Moore and M. O. Scully (Plenum, N. Y., 1986);
    E. Joos and H. D. Zeh, Z. Phys. B59, 223 (1985);
    A. O. Caldeira and A. J. Leggett, Phys. Rev. A31, 1059 (1985)
    W. G. Unruh and W. H. Zurek, Phys. Rev. D40, 1071 (1989);
    J. P. Paz, S. Habib and W. H. Zurek, Phys. Rev. D47, 488 (1993);
    W. H. Zurek, S. Habib and J. P. Paz, Phys. Rev. Lett. (1993)

\noindent [5] B. L. Hu, J. P. Paz and Y. Zhang, Phys. Rev. D45, 2843 (1992);
              D47, 1576 (1993).

\noindent [6] M. Gell-Mann and J. B. Hartle, in {\it Complexity, Entropy
             and the Physics of Information}, ed. W. Zurek,
             Vol. IX (Addison-Wesley, Reading, 1990);
             Phys. Rev. D47, 3345 (1993)
             R. Griffiths, J. Stat. Phys. 36, 219 (1984);
             R. Omnes, Rev. Mod. Phys. 64, 339 (1992);
             H. F. Dowker and J. J. Halliwell, Phys. Rev. D46, 1580 (1992);
             E. Calzetta and B. L. Hu, in {\it Directions in General
             Relativity}, Vol 2, ed. B. L. Hu and T. A. Jacobson
             (Cambridge University Press, Cambridge 1993)

\noindent [7] See, e.g., K. Huang, {\it Statistical Mechanics}, 2 ed. (Wiley,
    New York, 1987)

\noindent [8] A. Mann, M. Revzen, H.Umezawa and Y. Yamanaka, Phys. Lett.
              A140, 475 (1989)

\noindent [9] B. L. Hu and Yuhong Zhang, in {\it Proc. 3rd Drexel Conference
              on Quantum Nonintegrability}, ed. D. H. Feng et al,
              (Gordon and Breach, Philadelphia, 1993);
              B. L. Hu and Yuhong Zhang, "Uncertainty Relation for a
Quantum Open System" University of Maryland preprint umdpp 93-162 (1993)

\noindent [10] R. Rubin, J. Math. Phys. 1, 309 (1960);
              J. Schwinger, J. Math. Phys. 2, 407 (1961);
              G. W. Ford, M. Kac, P. Mazur, J. Math. Phys. 6, 504 (1963);
              H. Dekker, Phys. Rev. A16, 2116 (1977);
              V. Hakim and V. Ambegoakar, Phys. Rev. A36, 3509 (1985);
              F. Haake and R. Reibold, Phys. Rev. A32, 2462 (1985).

\noindent [11] R. P. Feynman and F. L. Vernon, Ann. Phys. 24, 118 (1963)

\noindent [12] A. O. Caldeira and A. J. Leggett, Physica 121A, 587 (1983);
    A. J. Leggett et al, Rev. Mod. Phys. 59, 1 (1987)

\noindent [13] H. Grabert, P. Schramm, and G. -L. Ingold,
                     Phys. Rep. 168, 115 (1988)

\noindent [14] See, e.g., B. L. Schumacher, Phys. Rep. 135, 317 (1986).
    For more recent work, see, e.g., Y. S. Kim and W. Zachary, eds
    {\it Squeezed State and  Uncertainty Relation} (NASA Publication, 1992)

\noindent [15] Although this decoherence time was not given explicitly,
 it could be derived from the results of [5] and Paz et al in [4].

\noindent [16] B. L. Hu, A. Raval and Y. Zhang, in preparation

\end